\documentclass[aps,prl,reprint,amsmath,amssymb,nofootinbib,nobibnotes]{revtex4-2}

\usepackage{graphicx}
\usepackage{aas_macros}
\usepackage{verbatim}
\usepackage{xcolor}

\preprint{APS/123-QED}

\usepackage[hidelinks]{hyperref}

\newcommand{\prlheadings}{}
\makeatletter
\ifdefined\prlheadings
  \let\prl@origpar\par
  \renewcommand{\section}{\@ifstar{\prl@sec}{\prl@sec}}
  \newcommand{\prl@sec}[2][]{%
    \refstepcounter{section}%
    \par\addvspace{1.5ex}%
    \noindent\textit{#2:}\ignorespaces
    \def\par{\let\par\prl@origpar\ignorespaces}}
  \renewcommand{\subsection}{\@ifstar{\prl@subsec}{\prl@subsec}}
  \newcommand{\prl@subsec}[2][]{\refstepcounter{subsection}}
\fi
\makeatother

\begin{document}

\title{Fast radio burst dispersion is an unbiased tracer of matter on large scales}

\author{Shion~Andrew$^{1,2}$}
\thanks{Co-first author}
\email{shiona@mit.edu}
\author{Haochen~Wang$^{1,2}$}
\thanks{Co-first author}
\email{hcwang96@mit.edu}
\author{Kiyoshi~Masui$^{1,2}$}
\author{Josh~Borrow$^3$}
\author{Calvin~Leung$^{4,5,6}$}
\author{Ryan~Raikman$^{1,2}$}
\author{Matthieu~Schaller$^{7,8}$}
\author{Joop~Schaye$^8$}
\author{James~M.~Sullivan$^9$}
\thanks{Brinson Prize Fellow}

\affiliation{$^1$ Department of Physics, Massachusetts Institute of Technology, 77 Massachusetts Avenue Cambridge, MA 02139, USA}
\affiliation{$^2$ MIT Kavli Institute for Astrophysics and Space Research, Massachusetts Institute of Technology, 77 Massachusetts Avenue Cambridge, MA 02139, USA}
\affiliation{$^3$ Department of Physics and Astronomy, University of Pennsylvania, 209 South 33rd Street, Philadelphia, PA 19104, USA}
\affiliation{$^4$ Department of Astronomy, University of California Berkeley, Berkeley, CA, 94720, USA}
\affiliation{$^5$ Department of Physics, University of California Berkeley, Berkeley, CA, 94720, USA}
\affiliation{$^6$ Miller Institute of Basic Research, University of California Berkeley, Berkeley, CA, 94720, USA}
\affiliation{$^7$ Lorentz Institute for Theoretical Physics, Leiden University, PO Box 9506, 2300 RA Leiden, the Netherlands}
\affiliation{$^8$ Leiden Observatory, Leiden University, PO Box 9513, 2300 RA Leiden, the Netherlands}
\affiliation{$^9$ Center for Theoretical Physics --- a Leinweber Institute, Massachusetts Institute of Technology, Cambridge, MA 02139, USA}

\date{\today}

\begin{abstract}
The dispersion of fast radio bursts (FRBs) measures the column density of
free electrons, tracing the diffuse ionized gas that
contains more than $90\%$ of all baryons. On linear scales the FRB dispersion 
field is an approximately unbiased tracer of the matter distribution---an idea 
long assumed in the FRB large-scale structure
literature and recently formalized by  \citet{2025arXiv251011022Z}. This follows from baryon-mass conservation, 
which forces the total baryon field to have unit linear bias, 
with dispersion inheriting this bias up to small corrections from the stellar and 
neutral-gas components. We show these corrections can be bounded at the percent level using existing galaxy and
 21 cm surveys, and confirm with the FLAMINGO hydrodynamical simulations
that the electron bias varies 
at the percent level across a wide range of feedback prescriptions. The
dispersion--galaxy cross-power spectrum at linear scales directly constrains $B_8 \equiv
\sigma_8(\Omega_b/0.05)^{1/2}$, a baryonic analog of $S_8$, 
independently of feedback physics. Because most
of the per-object variance in dispersion is cosmological signal rather
than noise, $\sim\!10^5$ localized FRBs can match the
statistical power of $\sim\!10^8$ weak-lensing galaxy shape
measurements. FRB dispersion thus joins weak lensing and redshift-space
distortions as a new unbiased tracer of matter on large scales.
\end{abstract}

\maketitle

\section{Introduction}
\label{sec:intro}

Large-scale structure---the statistical spatial distribution of matter in the
Universe---is one of the richest sources of information about the origin, composition, and evolution of the Universe. Measuring it
requires identifying a tracer of the underlying matter density, such as galaxy
abundance or the brightness of line emission, mapping it over cosmological
volumes, and characterizing its statistics. The power spectrum, which quantifies
the amplitude of density fluctuations as a function of scale, is the principal
statistic of interest. Cosmological models relate the observed power spectrum
to the contents, expansion history, and initial conditions of the Universe.

A central challenge in this program is tracer bias. A region that is $10\%$
overdense in matter, for instance, could be $20\%$ overdense in galaxies,
simply because the complicated physics of galaxy formation need not track the
matter density one-to-one. On large scales, where density perturbations are
small, this
mismatch reduces to a single proportionality constant $b_t$, the tracer bias,
relating the tracer overdensity ($\delta_t$) to the matter overdensity
($\delta$): $\delta_t = b_t\,\delta$~\footnote{In this work, we ignore complications 
from massive neutrinos, so the matter overdensity field $\delta$ only includes the 
contributions from cold dark matter and baryons.}. Because the evolution of tracers like
galaxies is a complicated astrophysical process, $b_t$ is difficult to predict
from first principles. The overall amplitude of the tracer auto-power spectrum (which scales as $b_t^2$) therefore becomes
a nuisance parameter, and cosmological constraints from tracer surveys must be
inferred primarily from the shape of the power spectrum.

Unbiased tracers of matter can overcome this limitation, but only two of such
tracers are known to date.
Weak gravitational lensing measures the coherent distortion in background galaxy
shapes and the CMB due to foreground mass concentrations, probing the gravitational
potential projected along the line of sight. Redshift-space distortions, in
the linear regime described by Kaiser~\cite{1987MNRAS.227....1K}, arise because
peculiar velocities sourced by the gravitational potential shift observed
redshifts. The resulting anisotropic signature in the clustering pattern has an
amplitude determined by the growth rate of structure, independently of tracer
bias.

Fast radio burst (FRB) dispersion has recently emerged as a probe of the
baryonic component of
matter~\cite{2014ApJ...780L..33M,2015PhRvL.115l1301M}. As radio pulses from
FRBs propagate through the intervening medium, free electrons along the line of
sight impart a frequency-dependent delay---the dispersion measure (DM)---that is
proportional to the integrated column density of ionized gas. Because about
$90\%$ of the baryons in the Universe reside in a diffuse, ionized state in the
intergalactic medium and circumgalactic halos~\cite{2020ARA&A..58..363P}, DM is sensitive to the dominant
reservoir of ordinary matter.
The sensitivity to ionized gas makes DM an excellent probe of galaxy 
feedback physics, and recent literature has explored DM clustering 
statistics to jointly constrain feedback and cosmological parameters
~\cite{2014ApJ...780L..33M,2015PhRvL.115l1301M,2019PhRvD.100j3532M,2026ApJ...998..109S,2026arXiv260212174W,2022MNRAS.512.1730S,wang2025}. 
However, we show that by exploiting FRB 
dispersion as an unbiased tracer of matter specifically at linear scales (i.e., $k \lesssim 0.1\,h/\mathrm{Mpc}$),
cosmology can be directly constrained from DM statistics 
without modeling or interpreting feedback physics. 

In the first proposal to use FRBs as a large-scale structure probe,
\citet{2015PhRvL.115l1301M} noted that because dispersion traces nearly all of
the baryons, and baryons are conserved in the Universe, the linear bias of the
DM field should be close to unity. This assumption has been adopted either
implicitly or explicitly in much of the subsequent FRB large-scale structure
literature. \citet{2025arXiv251011022Z} formalized the argument in the context
of using FRB dispersion to perform a large-scale test of General Relativity,
and further showed that departures from a bias of unity could be constrained by
external measurements or astrophysical models to sub-percent accuracy. Here we
argue that this idea is more broadly applicable: FRB dispersion joins weak
lensing and redshift-space distortions as a linear-scale unbiased tracer of
the matter distribution, and measurements of power spectra involving FRBs
can directly constrain cosmological parameters.

\section{Baryon Conservation and Dispersion Bias}
\label{sec:idea}

We start by showing that dispersion provides a tracer of large-scale
structure whose bias deviates from unity only weakly and can be
determined by either external measurements or astrophysical models. 
We first consider the total baryon field, which on large scales traces 
the total matter distribution with unit bias, $b_b = 1$. 
This follows from two facts. First, baryon mass is conserved: no Standard Model process creates or
destroys baryons, so the total comoving baryon density is 
fixed~\footnote{A negligible fraction of baryons may accrete into black
holes or change mass by converting to other baryon species.}. Second, on
large scales, baryons and dark matter
respond identically to gravity (i.e., the equivalence principle). 
Pressure forces redistribute baryons only over
a limited range, so on scales larger than the displacement of baryons relative
to dark matter (i.e.\ the Lagrangian displacement of baryons in the dark matter
frame, $k\sim0.1\, h/\textrm{Mpc}$) baryons are identically distributed to
the total matter.

FRB dispersion traces only the free electrons in diffuse ionized gas, which
accounts for roughly $90\%$ of all baryons~\cite{2020ARA&A..58..363P}.  Since
this dominant component comprises nearly all baryons, it inherits the unity
bias of the total baryon field to good approximation 
(i.e., gas is found to be clustered less strongly than the total matter by only a few percent \cite{refId0}).  The remaining
$\sim\!10\%$ of baryons, which are locked in stars or neutral gas (atomic and
molecular), are untraced by dispersion, but their contribution to the bias can
be corrected using complementary observations: stellar mass densities from
galaxy photometric surveys and neutral gas inventories from 21-cm and CO
emission.

We now formalize this idea, following \citet{2025arXiv251011022Z} but with
slightly different notation. We start with the baryon density field,
a function of spatial position and redshift,
and separate it into three components:
\begin{equation} \label{eq:tot_dense}
    \rho_b = \rho_d + \rho_* + \rho_n,
\end{equation}
where $\rho_b$, $\rho_d$, $\rho_*$, and $\rho_n$ represent the densities of
baryons, the diffuse ionized gas to which FRB dispersion is sensitive, stars,
and neutral gas (atomic and molecular), respectively.  There are some
subtleties as to precisely which baryons are included in $\rho_d$, which we
discuss in the end matter. Likewise, $\rho_*$ and $\rho_n$ could be further
subdivided, but here we use a broad categorization for simplicity.

Splitting each density field into its background and perturbation fields,
$\rho_t = \bar{\rho}_t(1 + \delta_t)$, and defining the mean density
fractions $f_t \equiv \bar{\rho}_t / \bar{\rho}_b$, we have
\begin{equation}
    \delta_b = f_d \delta_d + f_* \delta_* + f_n \delta_n.
\end{equation}
Assuming each component is a biased tracer of the total matter,
$\delta_t = b_t \delta_m$, and that the total baryon bias is unity, we
find
\begin{equation} \label{eq:fdb}
    f_d b_d = 1 - f_* b_* - f_n b_n.
\end{equation}
As we show in the cosmological constraints section below, observables
constructed from the DM field are proportional to $f_d b_d$; it is this
product that plays the role of the tracer bias for dispersion.

Eq.~(\ref{eq:fdb}) means that we need only bound the
small correction terms, not predict the bias itself. Crucially, the structure
of the equation means that \emph{$f_d b_d$ is inherently insensitive to
astrophysical uncertainties}, even if we choose to model it (as we show with
simulations below) rather than measure the corrections.

Constraints on the correction terms in Eq.~(\ref{eq:fdb}) come from a variety of sources. 
The stellar fraction $f_*~\approx~5$--$7\%$ is constrained by galaxy surveys and
stellar population
modeling~\cite{2018MNRAS.475.2891D,2020ApJ...893..111L,2004ApJ...616..643F};
the neutral fraction $f_n~\approx~3$--$4\%$ by 21-cm
surveys~\cite{2018MNRAS.477....2J,2025arXiv251119620C,2023ApJ...947...16A} and molecular gas
inventories~\cite{2020ApJ...902..111W,2020ARA&A..58..363P}. The bias factors
are order-unity and can be similarly constrained with external measurements:
since their prefactors are small, they need only be measured to $\approx~30\%$
precision in order for $f_d b_d$ to be constrained to $\approx~3\%$
precision with present data. This could be further improved with concerted effort. 
A more comprehensive overview of these constraints is presented in the end matter.

\section{Validation with Simulations}
\label{sec:validation}

We now test the claim that $f_d b_d$ is insensitive to astrophysical
modeling uncertainties using the FLAMINGO suite of cosmological hydrodynamical
simulations~\cite{2023MNRAS.526.4978S,2023MNRAS.526.6103K} run in a $(1\,\text{Gpc})^3$ box.
While Zhou \& Zhang~\cite{2025arXiv251011022Z} conducted a similar test using
Illustris and IllustrisTNG, FLAMINGO is particularly suited to the task
at hand: its feedback prescriptions were designed to approximately span the
range of astrophysical modeling uncertainties, and the simulations cover a large enough cosmic volume 
to sufficiently model both linear and non-linear processes simultaneously. 
The feedback models are calibrated to two key observables:
the $z=0$ galaxy stellar mass function (SMF), which constrains the
stellar content of haloes and ensures realistic galaxy clustering,
and the gas mass fractions in low-$z$ groups and clusters of galaxies,
which directly regulate how much baryonic matter is retained in or expelled
from massive haloes ~\cite{2023MNRAS.526.6103K}. In contrast to other hydrodynamical simulations such as IllustrisTNG, FLAMINGO suite also contains strong feedback variants 
that are observationally consistent with recent kinematic Sunyaev-Zel'dovich
measurements~\cite{2025MNRAS.540..143M,2025PhRvD.112h3509H}.

Using the total free-electron and total matter density fields
extracted from FLAMINGO at redshifts z = 0, 0.5, 1, and 2 (taking the
total electron density as a proxy for the diffuse component), we compute
$b_d$ and $f_d$ as a function of redshift
across all feedback variants. We use the power spectra described in Leung et al 
\cite{2025arXiv250919514L, 1998ApJ...499...20J,2020JOSS....5.2430B, borrow2021projectingsphparticlesadaptive} 
at a scale of $k=0.08 \text{ Mpc}^{-1}$ to extract the bias factor $b_d^2$ from the ratio of the electron and total power spectra. 
Other linear scales return similar bias factors and the results are not sensitive to the exact choice of linear scale.
$f_d$ is obtained by inverting 
\begin{equation}
    \label{eqn:n_e_to_f_d}
    \bar{n}_e(z) = \Omega_b \rho_{\rm cr}\,(1+z)^3\, f_d(z)\, m_p^{-1} ( 1 - Y/2 ),
\end{equation}
where $\bar{n}_e(z)$ is the mean proper free-electron number density (obtained from simulation snapshots), 
$\Omega_b$ is the present-day mass fraction of baryons,
$\rho_{\rm cr}\equiv 3 H_0^2/(8\pi G)$ is the present-day critical density, $m_p$ the proton
mass, and $Y$ the helium mass fraction~\cite{2020Natur.581..391M}.

\begin{figure*}
    \includegraphics[width=\textwidth]{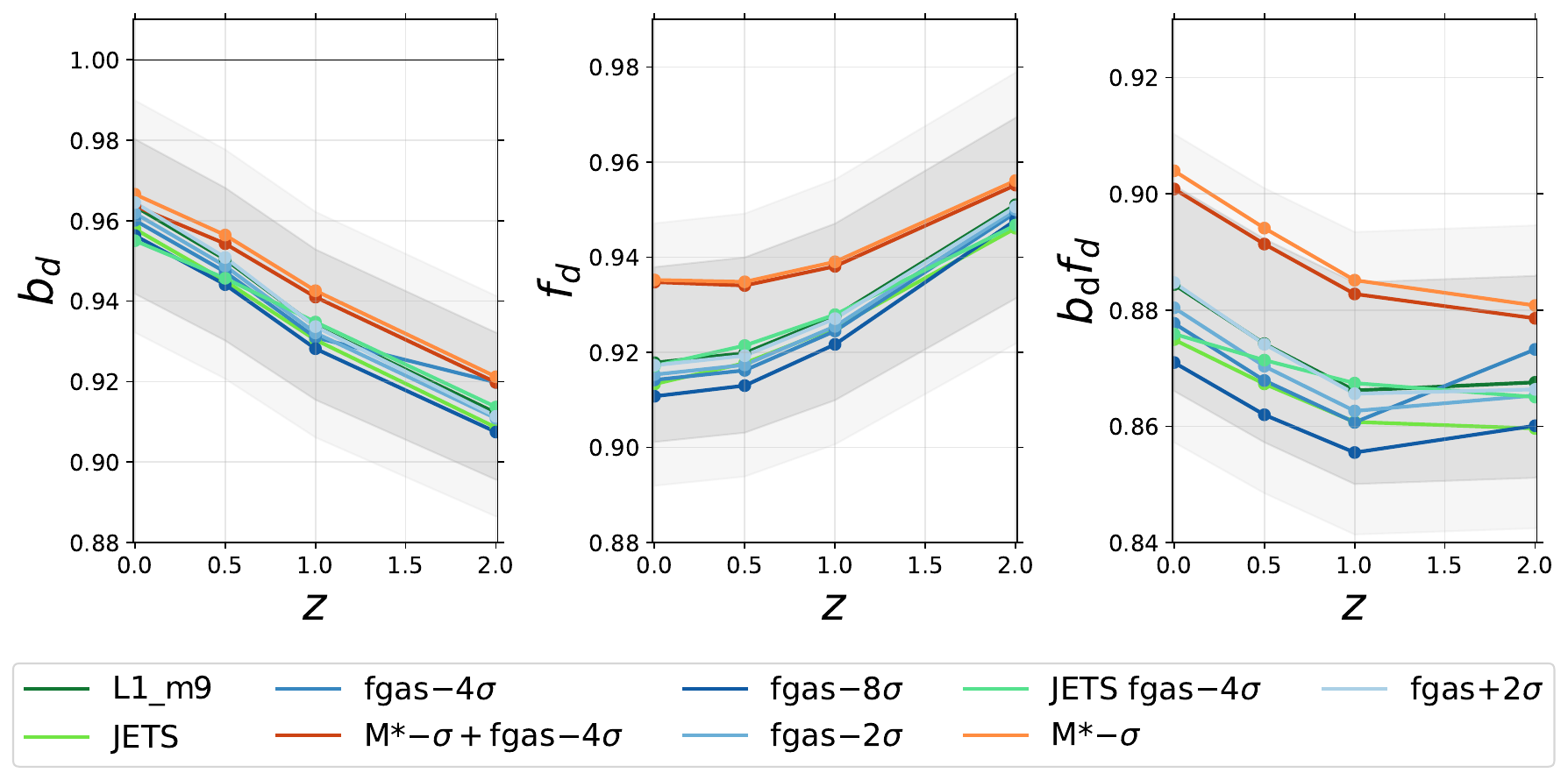}
    \caption{Diffuse baryon bias $b_d$ (left), diffuse baryon fraction
    $f_d$ (middle), and their product (right) as a function of redshift
    for FLAMINGO feedback variants at a fixed cosmology described 
    in \cite{2023MNRAS.526.4978S}. Dark (light) gray bands show $\pm 2\%~(3\%)$ around the
    fiducial feedback model (labeled ``L1\_m9") at each redshift. Each variant is a simulation whose feedback
    was recalibrated so that the halo gas fraction (\texttt{fgas}) or
    stellar mass function (\texttt{M*}) deviates from the observed values
    by the indicated number of standard deviations;
    \texttt{JETS} replaces thermal AGN feedback with directed kinetic
    jets, and \texttt{L1\_m9} is the fiducial run calibrated to match
    both observables. See Table~8 and
    Section~6.4 of \cite{2023MNRAS.526.6103K} for details. The large-scale distribution of diffuse ionized gas
    is largely insensitive to astrophysical modeling uncertainties: at fixed
    redshift, $b_d$ and $f_d$ individually vary by 
    $\approx~2\%$ across models, while their product varies by
    $\approx~3\%$. The corresponding values for different cosmological variants (not shown) vary at the percent level and is subdominant to the
    variation due to uncertainty in feedback prescriptions.}
    \label{fig:bias_vs_redshift}
\end{figure*}

Fig.~\ref{fig:bias_vs_redshift} shows that $b_d$ and $f_d$
individually vary by no more than $\approx~2\%$ across models at any given
redshift, while their product varies by $\approx~3\%$, confirming the 
expectation that on large scales, the distribution of diffuse ionized baryons
is insensitive to astrophysical feedback processes. Note that as the simulations were 
calibrated against the stellar mass function, uncertainties in the underlying measurements (discussed in the end matter) 
are carried into the normalization for $f_*$, potentially leading to a shared, systematic, offset 
in the absolute value of $f_d b_d$. The key
result here is the small \emph{spread} across feedback variants, despite wide 
variability in the feedback and resulting non-linear effects, which
demonstrates that astrophysical modeling uncertainties contribute at most
at the percent level.


\section{Cosmological Parameters from Dispersion Statistics}
\label{sec:power_spectra}

Having established that the dispersion bias $f_d b_d$ is close to unity
and insensitive to astrophysical uncertainties, we now ask what
cosmological information can be extracted from large-scale statistics of
the DM field.  We show that cross-correlating dispersion with a galaxy
survey yields a measurement of
$B_8 \equiv \sigma_8(\Omega_b/0.05)^{1/2}$, the baryonic analog of
the parameter $S_8$ to which weak lensing is most sensitive \cite{descollaboration2026darkenergysurveyyear, 2023PhRvD.108l3519D,2025A&A...703A.158W}.

The FRB DM field along an observed direction $\hat{n}$ out to redshift $z$
is
\begin{align} \label{eq:d_field}
    D(\bm{\hat{n}}, z) = \int_0^{\chi(z)} \frac{d\chi'}{(1+z')^2}\,\bar{n}_e(z')\,[1+\delta_d(\bm{\hat{n}}, \chi')],
\end{align}
where $\chi(z)$ is
the comoving distance, and the two powers of $(1+z)^{-1}$ convert proper path
length and dispersive delay to the observer frame.
The mean electron density $\bar{n}_e(z)$ is given by Eq.~(\ref{eqn:n_e_to_f_d}).

Several clustering statistics involving FRB dispersion have been
considered, differing in how the DM measurement is used, whether FRB
host galaxies and redshifts are known, and whether the statistic is an
auto-correlation or a cross-correlation with other
tracers~\cite{2014ApJ...780L..33M,2015PhRvL.115l1301M,2019PhRvD.100j3532M,2026ApJ...998..109S,2026arXiv260212174W,2022MNRAS.512.1730S, wang2025}.
To illustrate the cosmological parameter dependence, we consider the
case where dispersion measures of background FRBs with known redshifts
are cross-correlated with foreground galaxy number counts, as first
proposed by \citet{2019PhRvD.100j3532M}. This is the mathematically
simplest statistic, as the fixed galaxy redshift eliminates redshift
integrals. It also does not suffer from contamination by clustering of
FRB sources and is only weakly sensitive to selection
effects~\cite{2026ApJ...998..252C}. Other statistics share the same principal parameter
dependencies.

The DM--galaxy angular cross-power spectrum for galaxies in a thin shell at redshift $z_g$
and a DM field from background FRBs is~\cite{2019PhRvD.100j3532M}
\begin{equation} \label{eq:cldg}
    C_l^{Dg} = \frac{\bar{n}_e(z_g)}{(1 + z_g)^2\,\chi_g^2} P_{dg}(k = l/\chi_g, z_g),
\end{equation}
where $l$ is the sky multipole and $P_{dg}(k, z)$ is the diffuse-gas--galaxy
power spectrum at spatial wave number $k$.
We use the linear bias model to relate power spectra to the matter power
spectrum, $P_{dg}(k) = b_d b_g P(k)$, and define the amplitude-normalized
power-spectrum $\tilde{P}(k) \equiv P(k) / \sigma_8^2$, where the $\sigma_8$
parameter captures the power-spectrum amplitude information. With these
definitions, and ignoring well-measured constants, we have
\begin{align}
    C_l^{Dg} &\propto \Omega_b\,\sigma_8^2\,
        f_d b_d\, b_g\, \frac{(1+z_g)\,H_0^2}{\chi_g^2}\,\tilde{P}_m(l/\chi_g, z_g).
        \label{eq:cldg_full}
\end{align}

It is illustrative to compare this to the analogous statistic in weak
gravitational lensing. Similarly ignoring well-measured constants, the
galaxy--galaxy lensing angular cross-power spectrum is
\begin{align}
    C_l^{\gamma g} &\propto \Omega_m\,\sigma_8^2\,
        b_g\, \frac{(1+z_g)\,H_0^2}{\chi_g^2}\,g(z_s, z_g)\,\tilde{P}_m(l/\chi_g, z_g),
        \label{eq:clgammag}
\end{align}
where $z_s$ is the lensing source galaxy redshift, and $g(z_s, z_g) \equiv \bigl[H(z_g)\,\chi_g\bigr]\,(\chi_s - \chi_g)/\chi_s$
is a dimensionless geometric factor that depends only weakly on cosmological
parameters.

Comparing Eqs.~(\ref{eq:cldg_full}) and~(\ref{eq:clgammag}),
dispersion and shear have nearly identical parameter dependencies:
dispersion has $\Omega_b$ in place of lensing's $\Omega_m$, plus the
factor $f_d b_d$ which can be separately constrained to a few percent.
Although the power spectrum shape $\tilde{P}_m(l/\chi_g, z_g)$
is dependent on $\Omega_m$ rather than $\Omega_b$, the ratio $\Omega_b/\Omega_m = 0.157$ is well measured 
from the CMB to within a 1\% uncertainty \cite{2020A&A...641A...6P} and does not evolve 
due to the nearly perfect mass conservation of baryons.
We can then treat the remaining $\Omega_m$ dependence in Eq.~(\ref{eq:cldg_full}) as $\Omega_b$ dependence, 
resulting in the dominant parameter combination 
\begin{equation}
    B_8 \equiv \sigma_8 \left( \frac{\Omega_b}{0.05} \right)^{0.5},
\end{equation}
in analogy to $S_8 \equiv \sigma_8 (\Omega_m/{0.3})^{0.5}$ 
that cosmic shear analysis is mostly sensitive to~\footnote{The power of 0.5 on $\Omega_m$ mostly comes from 
the matter power spectrum shape dependence in the nonlinear regime. At linear scales, 
the lensing power is approximately proportional to $\Omega_m^{0.7}$ according to \cite{1997ApJ...484..560J}. We could alternatively have defined 
$B_8 \equiv \sigma_8 (\Omega_b/0.05)^{0.7}$. Given that the difference between 0.5 and 0.7 is small and that 
the power-law fit to the lensing signal is only approximate, 
we keep the power of 0.5 on $\Omega_b$ in this paper to follow the convention of the lensing literature.}~\cite{1997ApJ...484..560J}.



Intuitively, lensing shear and the DM field are both line-of-sight
integrals over unbiased tracers of matter: the former is sourced by the
total matter density ($\Omega_m$), the latter by baryonic matter
($\Omega_b$). Combining a late-time measurement of $B_8$ with an
early-Universe determination of $\Omega_b h^2$ from the CMB or
big-bang nucleosynthesis breaks the degeneracy between $\sigma_8$ and
$\Omega_b$, yielding an independent measurement of $\sigma_8$.

For mathematical simplicity we have focused on cross-correlations with
foreground galaxies, which introduces the galaxy bias $b_g$ as a nuisance
parameter for both lensing and dispersion. In practice, $b_g$ is constrained
by jointly fitting the galaxy auto-power spectrum; the sheer number of galaxies in present
redshift surveys makes this unlikely to be a limiting factor \citep[e.g.][]{2026AJ....171..285D, 2025JCAP...09..008A}. More broadly, we expect
all clustering statistics involving dispersion to share the primary
cosmological parameter dependence on $B_8$, including DM auto-correlations and
cross-correlations with other tracers such as weak lensing.

The comparison with lensing also highlights a key advantage.
For sources at $z = 1$, the weak-lensing shear signal power at
$\ell = 100$---corresponding to
$k \approx 0.1\;h\;\mathrm{Mpc}^{-1}$ at the midpoint of the line of
sight, close to the onset of nonlinearity---is
$C_\ell^{\gamma\gamma} \approx 10^{-8}$ (the shear and dispersion spectra share
a similar shape, so the comparison below does not depend on the chosen
$\ell$). The shape noise power spectrum is
\begin{equation}
  N_\ell^{\gamma} = \frac{\sigma_\epsilon^2}{\bar{n}_g},
\end{equation}
where $\sigma_\epsilon \approx 0.3$ is the intrinsic ellipticity
dispersion and $\bar{n}_g$ is the areal density of source galaxies.
For FRB dispersion at the same $\ell$ and source redshift, the signal
power is
$C_\ell^{DD} \approx 1\;(\mathrm{pc\;cm^{-3}})^2$~\cite{2019PhRvD.100j3532M},
while the noise power spectrum is
\begin{equation}
  N_\ell^{D} = \frac{\sigma_D^2}{\bar{n}_f},
\end{equation}
where $\sigma_D \approx 100\;\mathrm{pc\;cm^{-3}}$ is the host-galaxy
DM spread and $\bar{n}_f$ is the areal density of localized FRBs.
The ratio of signal power to noise power for the two probes is equal when
$\bar{n}_f \approx 10^{-3}\,\bar{n}_g$: for surveys with equal sky areas,
FRB dispersion requires
roughly a thousandth the sources as cosmic shear to achieve the
same sensitivity.

This comparison is restricted to linear scales.
Cosmic shear is an unbiased tracer of the total matter at \emph{all} scales, and in practice most of its
statistical power comes from nonlinear scales ($\ell \gtrsim$ a few hundred)
where measurement errors are smallest. On those scales, the DM bias departs
from unity and becomes sensitive to baryonic physics, so the simple
equal-footing comparison above no longer applies. However, the sensitivity
of DM to baryonic feedback on small scales is itself a powerful asset:
cross-correlating dispersion with lensing can isolate and remove the
feedback signal that is the dominant astrophysical systematic in
weak-lensing cosmology~\cite{2025arXiv250919514L}.

We note that measurements of power spectra amplitudes are generically more
sensitive to systematic errors than those of power spectra shapes.
DM-dependent selection
effects~\cite{2026ApJ...998..252C}, catastrophic redshift errors from
host
misidentification~\cite{2024ApJ...977L...4H,2025arXiv250620774M},
and the construction of unbiased power spectrum
estimators~\cite{2001PhRvD..64f3001T} all require careful treatment; a
detailed discussion is given in the end matter.

\section{Conclusions}
\label{sec:conclusions}

We have argued that fast radio burst dispersion is an unbiased tracer of
the large-scale matter distribution. The argument rests on baryon
conservation: because no astrophysical process creates or destroys baryons,
and because gravity is the only force relevant on large scales, the total
baryon field traces the matter field with unit bias. Since the diffuse
ionized gas probed by FRB dispersion accounts for roughly $90\%$ of all
baryons, its bias departs from unity by only the small corrections due to
stellar and neutral-gas components---corrections that can be bounded at
the percent level from existing observations or marginalized over in a
joint analysis.

This property elevates FRB dispersion from an astrophysical diagnostic to
a cosmological tool. Dispersion clustering statistics will be sensitive to
$B_8 \equiv \sigma_8(\Omega_b/0.05)^{1/2}$, providing a measurement of
the amplitude of matter fluctuations that is independent of, and
complementary to, weak-lensing determinations of $S_8$. The favorable
noise properties of the DM field---where the cosmological signal
constitutes a substantial fraction of the per-object variance, in contrast
to the shape-noise-dominated regime of cosmic shear---mean that
$\sim\!10^5$ localized FRBs with known redshifts could achieve
statistical power comparable to current lensing surveys with
$10^8$ galaxy shapes. This comparison applies on linear scales.
Lensing retains its advantage on nonlinear scales where most of its
statistical power resides.

Realizing these measurements will require careful control of amplitude
systematics, including redshift and localization errors, the distinction
between electron and baryon mass fractions, and the construction of
unbiased power spectrum estimators. None of these challenges appears
insurmountable, and several are shared with---and can draw on experience
from---existing galaxy and lensing surveys. With FRB detection rates
growing rapidly and localization capabilities expanding through
instruments such as CHORD~\cite{2019clrp.2020...28V} and the
DSA~\cite{2019BAAS...51g.255H}, dispersion-based cosmology is poised to
deliver competitive constraints on the growth of structure.

\section{Acknowledgments}
\label{sec:acknowledgments}
This work used the DiRAC@Durham facility managed by the Institute for Computational Cosmology 
on behalf of the STFC DiRAC HPC Facility (www.dirac.ac.uk). 
The equipment was funded by BEIS capital funding via STFC capital 
grants ST/K00042X/1, ST/P002293/1, ST/R002371/1 and ST/S002502/1, Durham University and 
STFC operations grant ST/R000832/1. DiRAC is part of the National e-Infrastructure.
K. W. M. holds the Adam J. Burgasser Chair in Astrophysics and is supported by an NSF Grant (2008031). 
C. L. acknowledges support from the Miller Institute for Basic Research at UC Berkeley.
J. M. S. acknowledges that support for this work was provided by The Brinson Foundation through a Brinson Prize. 
The scientific ideas and arguments presented in this work originated with the
authors. Claude (Anthropic) was used extensively as a writing and research
tool throughout the preparation of this manuscript, including drafting and
editing prose, performing literature searches, and managing references.

\bibliography{references}

\appendix

\section{The Baryon Correction Terms}
\label{sec:endmatter_constraints}

Here we survey the present observational constraints on the correction
terms $f_* b_*$ and $f_n b_n$ appearing in Eq.~(\ref{eq:fdb}), which
determine the dispersion bias $f_d b_d$ discussed above.

The largest contribution to the correction terms is the stellar component.
Multi-wavelength galaxy surveys place living stars at $\approx~3$--$5\%$ of
all baryons at $z~\approx~0$~\cite{2018MNRAS.475.2891D}, with a realistic
uncertainty of $\approx~40\%$ driven by the assumed initial mass function,
details of SED fitting~\cite{2020ApJ...893..111L}, and the treatment of
diffuse stellar light. Including stellar remnants and brown
dwarfs~\cite{2004ApJ...616..643F} brings the total to $f_*~\approx~
5$--$7\%$.

The stellar bias $b_*$ can in principle be measured from the power spectrum of
stellar-mass-weighted galaxies, inferred
from the stellar mass function through abundance matching, or modelled from
simulations that match galaxy clustering as a function of mass, but to our knowledge no dedicated estimate exists in the
literature. Since $f_*$ is small, even an order-unity uncertainty in
$b_*$ shifts $f_d b_d$ by only a few percent.

The neutral term $f_n b_n$ encompasses atomic hydrogen (HI), its associated
neutral helium, and molecular gas (primarily H$_2$). The dominant component is
HI\@. At $z~\approx~0$, the HI mass function is well measured by 21~cm galaxy
surveys such as ALFALFA~\cite{2018MNRAS.477....2J}. At higher redshift,
21~cm intensity mapping measures a signal proportional to
$f_\mathrm{HI} b_\mathrm{HI}$---the combination that enters
Eq.~(\ref{eq:fdb}) directly. Recent CHIME auto- and cross-power spectrum
detections at $z~\approx~1$~\cite{2025arXiv251119620C,2023ApJ...947...16A}
constrain this product to $\approx~25\%$ precision, with a central value of
$f_\mathrm{HI} b_\mathrm{HI}~\approx~0.025$ at $z~\approx~1$.
Neutral helium associated with HI regions adds $\approx~1/3$ by mass (from
the primordial helium fraction $Y_p~\approx~0.245$; see
\cite{2025arXiv251011022Z} for an explicit treatment).
Molecular gas adds a further subdominant contribution: at $z = 0$ the
molecular fraction is $\approx~0.3\%$ of all baryons, rising to $\approx~1\%$
near the peak of cosmic star formation at $z~\approx~
1$--$2$~\cite{2020ApJ...902..111W,2020ARA&A..58..363P}, and can be
constrained by CO intensity mapping or empirical scaling relations.
Including all components, the total neutral correction is $f_n b_n~\approx~
0.03$--$0.04$, depending on redshift. Because this entire term contributes
at only the few-percent level in Eq.~(\ref{eq:fdb}), even approximate knowledge
of its constituents is sufficient: an order-unity determination of $f_n
b_n$ shifts $f_d b_d$ at only the percent level.

Remaining baryon reservoirs, such as condensed matter (dust and planets), 
supermassive black holes, and metals, contribute negligibly to the baryon
budget~\cite{2004ApJ...616..643F}.

\section{Systematic Errors}
\label{sec:endmatter_systematics}
\subsection{Dispersive baryons}
A subtlety in the baryon accounting above is which ionized baryons actually
contribute to the
dispersion of observed FRBs. The warm ionized medium (WIM) of the ISM,
and to a lesser extent H\textsc{ii} regions, are by definition part of the
diffuse ionized component $\rho_d$. By mass, the WIM accounts for roughly
$0.5\%$ of all baryons at $z \approx 0$~\cite{2009RvMP...81..969H}, though this fraction is highly uncertain
as it is difficult to observe outside the Milky Way. While this gas does cause dispersion, intersections of FRB sightlines with the
ISM of intervening galaxies are very rare given their small geometric
covering fraction~\cite{2018MNRAS.474..318P}---placing this deep in the
non-Gaussian statistics regime. Furthermore, FRBs whose
sightlines do intersect such dense ionized gas are expected to be
strongly scattered and temporally
broadened~\cite{2022ApJ...934...71O,2018ApJ...863...48C}, pushing them
below the detection threshold of flux-limited surveys; the geometric
lever arm between source and scattering screen amplifies the temporal
broadening for intervening galaxies.
Understanding this effect will require
further study, but its impact on $f_d b_d$ is expected to be small.

Finally, Eq.~(\ref{eqn:n_e_to_f_d}) accounts only for hydrogen and
helium. Heavier elements yield fewer free electrons per unit baryon
mass---and fewer still in practice, since inner-shell electrons remain
bound at IGM temperatures---but metals dissolved in the diffuse IGM and
CGM amount to only $\sim\!0.01$--$0.1\%$ of all baryons by
mass~\cite{2014MNRAS.438..262P,2020ARA&A..58..363P}, making this
composition correction negligible.

\subsection{Power Spectrum Amplitude Errors}
\label{sec:endmatter_ps_amplitude}
Measurements of power spectra amplitudes are generically more prone to
systematics than those of power spectra shapes.
Cross-correlation statistics such as $C_\ell^{Dg}$ are naturally robust to
host-DM contributions, which add noise but do not bias the measurement since
they are uncorrelated with the foreground large-scale structure.
DM-dependent selection effects intrinsic to the FRB survey---such as
incompleteness at high DM due to dispersive smearing---can bias the
inferred power spectrum amplitude, but
operate predominantly on small, nonlinear
scales~\cite{2026ApJ...998..252C}.
A second concern is catastrophic redshift errors from misidentified host
galaxies. For instance, FRBs hosted by faint or low-mass dwarf
galaxies~\cite{2024ApJ...977L...4H,2025arXiv250620774M} may lack identifiable
hosts in the survey catalog, leading to
misassociation with a brighter neighbor. If a foreground FRB is assigned to the wrong host and placed in the
\emph{background} of the galaxy sample, its DM fluctuation will be uncorrelated
with the foreground galaxies and therefore suppress the
measured cross-power spectrum amplitude. An analysis by Andersen
et al. (in prep) shows
that for a CHIME-like FRB survey ($z \leq 2.5$) with subarcsecond-localized FRBs and 
sufficient optical followup imaging depth (down to $r$-band magnitude $\approx~24$),
the misassociation rate is generally expected to be $\lesssim$ 1\%. 
Nevertheless, this systematic should be considered with care, since 
the exact misassociation rate depends sensitively on the 
localization precision of the FRB survey, 
the optical imaging depth, 
and the details of the host association framework employed (for instance, the
assumed offset distribution of FRBs from their host).
If using photometric redshifts for either the galaxy tracers or the FRB hosts,
catastrophic redshift errors could have a similar effect.

Beyond astrophysical and observational systematics, the power spectrum
estimator itself must not introduce a
multiplicative bias on the amplitude. One can choose an existing unbiased power spectrum estimator 
for the FRB context \citep[e.g.][]{Alonso_2019, Baleato_Lizancos_2024, Wolz_2025} or 
forward-model the estimator's amplitude response. For both the galaxy and FRB fields,
the survey window function---set by the footprint, incompleteness, and
radial selection---must be modeled accurately, typically through random
catalogs or simulation-based mocks. Errors in these propagate directly
into the inferred power spectrum amplitude, making accurate
forward-modeling of both galaxy and FRB detectability
as a function of position and
redshift essential.


\end{document}